\documentclass[a4paper,fleqn]{cas-dc}

\usepackage[authoryear,longnamesfirst]{natbib}

\usepackage{url}
\usepackage{lscape} 
\usepackage{longtable}
\usepackage{booktabs}
\usepackage[figuresright]{rotating}
\usepackage{tabularx}

\usepackage{hyperref} 
\hypersetup{
  colorlinks=true,
  citecolor=cyan,
  linkcolor=cyan,
  urlcolor=cyan}

\def\tsc#1{\csdef{#1}{\textsc{\lowercase{#1}}\xspace}}
\tsc{WGM}
\tsc{QE}
\tsc{EP}
\tsc{PMS}
\tsc{BEC}
\tsc{DE}

\begin{document}
\let\WriteBookmarks\relax
\def\floatpagepagefraction{1}
\def\textpagefraction{.001}
\let\printorcid\relax

\shorttitle{Towards Human-AI Collaborative Urban Science Research Enabled by Pre-trained Large Language Models}

\shortauthors{Jiayi Fu et~al.}

\title [mode = title]{Towards Human-AI Collaborative Urban Science Research Enabled by Pre-trained Large Language Models}



%
\author[1]{Jiayi Fu}
                        




\credit{Conceptualization of this study, Methodology, Data curation, Writing - Original draft preparation, Software}

\author[1]{Haoying Han}
\credit{Data curation, Revising the draft}
\cormark[1]
\ead{hanhaoying@zju.edu.cn}
\author[1]{Xing Su}
\credit{Data curation, Revising the draft}

\affiliation[1]{organization={College of Civil Engineering and Architecture, Zhejiang University},
    city={Hangzhou},
    postcode={310058}, 
    country={China}}


\author[2]{Chao Fan}
\cormark[2]
\ead{cfan@g.clemson.edu}
\credit{Conceptualization of this study, Methodology, Writing \& Revising - Original draft preparation}

\affiliation[2]{organization={School of Civil and Environmental Engineering and Earth Sciences, Clemson University},
    city={Clemson},
    state={SC},
    postcode={29634},
    country={USA}}




\begin{abstract}
Pre-trained large language models (PLMs) have the potential to support urban science research through content creation, information extraction, assisted programming, text classification, and other technical advances. In this research, we explored the opportunities, challenges, and prospects of PLMs in urban science research. Specifically, we discussed potential applications of PLMs to urban institution, urban space, urban information, and citizen behaviors research through seven examples using ChatGPT. We also examined the challenges of PLMs in urban science research from both technical and social perspectives. The prospects of the application of PLMs in urban science research were then proposed. We found that PLMs can effectively aid in understanding complex concepts in urban science, facilitate urban spatial form identification, assist in disaster monitoring, and sense public sentiment. At the same time, however, the applications of PLMs in urban science research face evident threats, such as technical limitations, security, privacy, and social bias. The development of fundamental models based on domain knowledge and human-AI collaboration may help improve PLMs to support urban science research in future.

\end{abstract}




\maketitle

\section{Introduction}
As the most intricate creation of humankind, cities are convoluted systems comprised of multiple dimensions and factors. Consequently, urban research has evolved into a complex and significant social undertaking \citep{emmi2008urban, marshall2012planning}. Furthermore, the technological revolution, the proliferation of big data in cities, and the dissemination of artificial intelligence have not only transformed cities but have also altered the manner in which urban researchers investigate them \citep{wang2023defining} technologies such as Machine Learning (ML), Deep Learning (DL), and their applications in Natural Language Processing (NLP) and Computer Vision (CV) have gained extensive usage in the realm of urban science research \citep{cai2021natural, casali2022machine, wang2022unsupervised}. These emerging technologies pose an opportunity to traditional urban research methodologies and propel urban research towards a quantitative, computational and intelligent direction. However, despite their promising potential, several obstacles hinder their applications, such as low robust performance \citep{goodfellow2014explaining}, algorithmic and technical constraints \citep{cai2021natural}, and insufficient semantic comprehension \citep{bender2020climbing}. Whether these issues can be resolved via novel technologies or tools constitutes a topic worthy of examination in current urban science research.

Pre-trained large language models (PLMs), such as ChatGPT \citep{IntroducingChatGPT}, have the potential to play a pivotal role in tackling these challenges. PLMs are a new paradigm of NLP \citep{li2021pretrained} that can be pre-trained on large-scale text corpora using self-supervised learning to simplify various complex natural language processing issues into straightforward fine-tuning problems \citep{qiu2020pre}. At present, PLMs have evolved monolingual such as BERT \citep{devlin2018bert}, GPT \citep{openai2023gpt4}, and multilingual training models such as mBERT \citep{devlin2018bert}, XLM-R \citep{conneau2019unsupervised}. One notable model among them is ChatGPT, a large language model that has been trained in PLMs based on autoregressive language \citep{else_abstracts_2023, yang2023harnessing}. By integrating various cutting-edge techniques such as unsupervised learning, and instruction fine-tuning \citep{wu2023brief}, ChatGPT boasts formidable content generation capabilities through its artificial intelligence generated content (AIGC) technology. This capability enables PLMs to independently learn from data and produce sophisticated and seemingly intelligent outcomes \citep{van2023chatgpt}. Due to its exceptional proficiency in text learning, text classification, information extraction, and text generation \citep{owens2023nature, wang2023chatgpt, wang2023chat}, PLMs have demonstrated its immense potential in diverse fields including finance \citep{dowling2023chatgpt}, medicine \citep{biswas2023role, jungwirth2023artificial, verhoeven2023chatgpt}, education \citep{kooli2023chatbots, yang2023use}, and environment \citep{an2023chatgpt, biswas2023potential, zhu2023chatgpt}.

\begin{figure}
	\centering
		\includegraphics[width=\linewidth, scale=0.75]{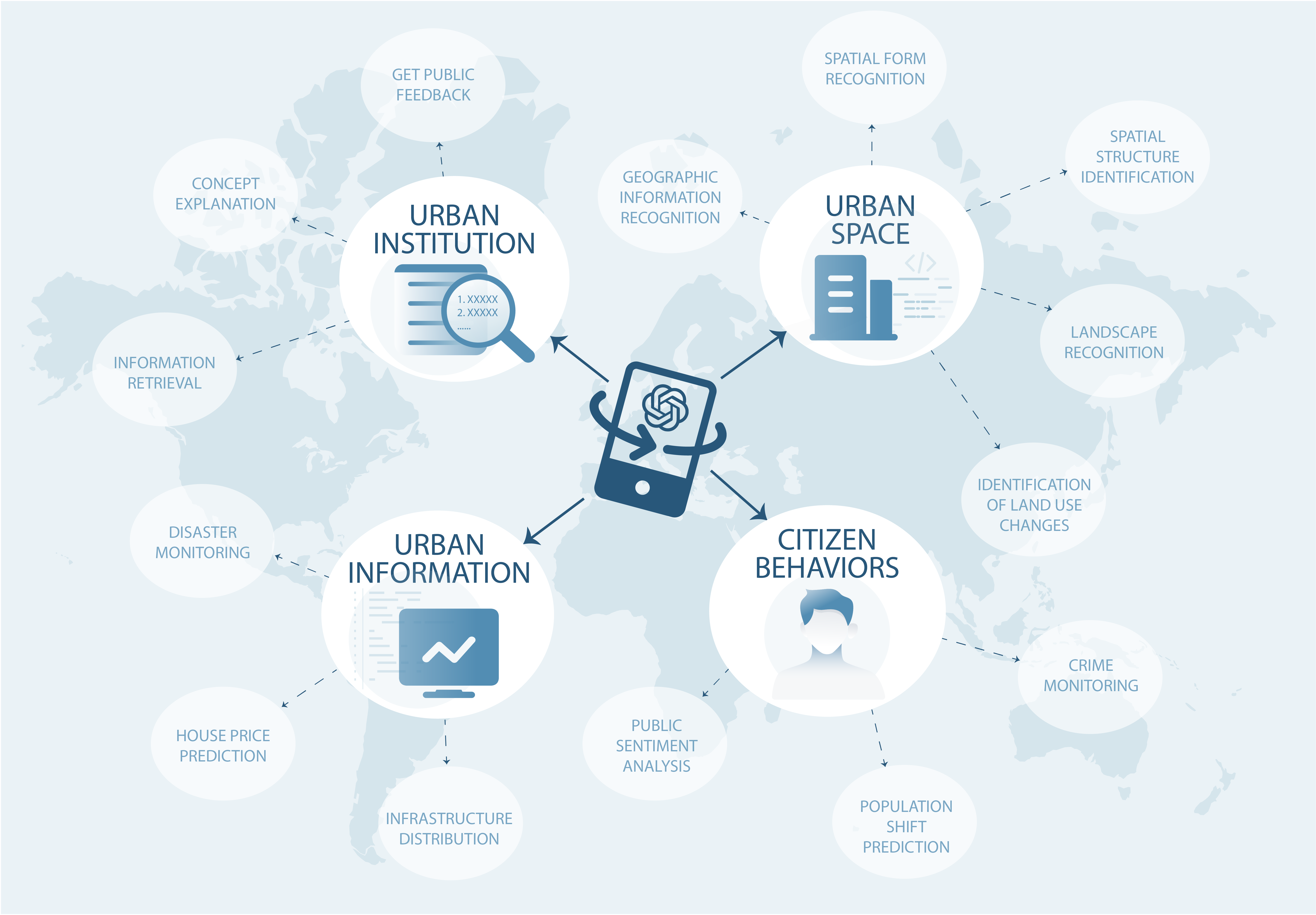}
	\caption{PLMs in urban science research}
	\label{FIG:02}
\end{figure}

PLMs are expected to play a crucial role in advancing urban research through various means, such as simplifying the interpretation of complex urban concepts, automating repetitive tasks programming on analyzing urban data, and improving the utilization of multi-disciplinary knowledge for urban science research (see Figure \ref{FIG:02}). PLMs will become a potent tool to support urban researchers in their efforts to achieve a new level of depth in urban research. In this paper, we utilize ChatGPT as a tool to investigate the opportunities and challenges associated with PLMs in urban research. The structure of this paper is as follows: In the second section, we outline the possible contributions of PLMs in urban institution, urban space, urban information, and citizen behaviors. We then examine potential issues and challenges facing PLMs in urban research from both technical and social perspectives. Finally, we explore possible directions for PLMs in future urban research.

\section{Opportunities}
\subsection{Urban institution}
The domain of urban institutional research comprises a wide range of topics, including but not limited to institutional design, public policy development, public comprehension of policy, and public policy response \citep{farazmand2023global}. This involves handling a significant amount of text-based data. As AI models, PLMs offer superior intelligent question and answer, text classification, and text generation capabilities. PLMs can comprehend the queries or questions of the researcher and respond with accurate and lucid language in both restricted or open domain Q\&A \citep{wu2023brief}. PLMs can also extract crucial information from a city system document to provide a summary of the main content of the document \citep{min2021recent}. Moreover, text classification is a distinct advantage of PLMs. PLMs can tell the positive and negative sentiment of texts, which aid researchers in obtaining prompt public feedback on urban institutions, discerning the public's key requirements for institutions or policies \citep{karduni2020anatomy}, and enables policymakers to comprehend the underlying reasons for public endorsement or opposition to urban institutions \citep{luo2019frontiers}. This utilization of public opinion helps to advance the construction of urban institutions. We will demonstrate the potential of PLMs in urban institution research through two examples. 

\begin{figure}
	\centering
		\includegraphics[width=\linewidth, scale=0.75]{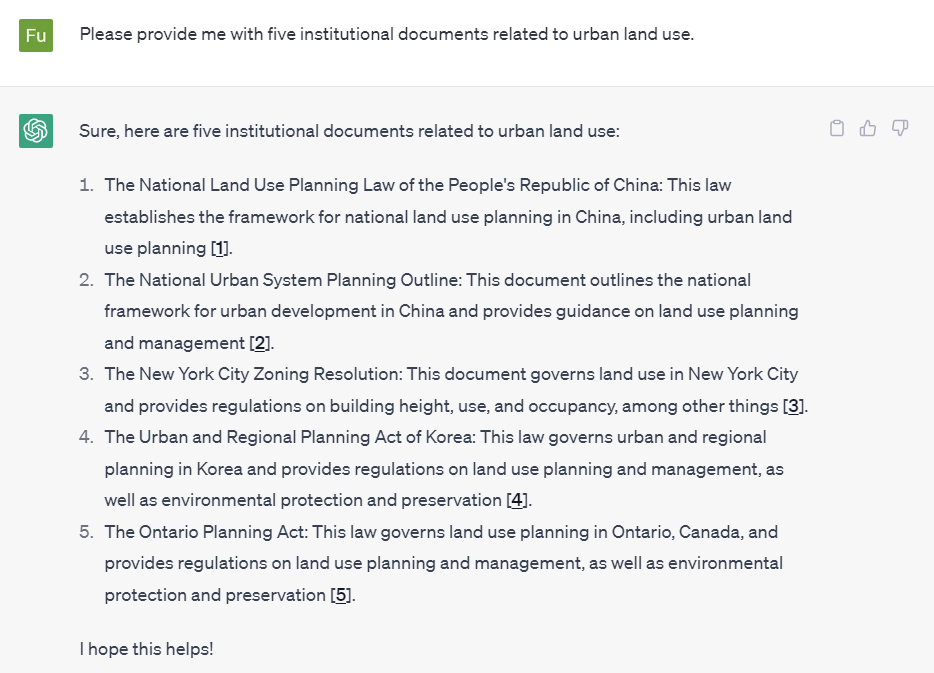}
	\caption{ChatGPT for searching documents (generated on April 22, 2023)}
	\label{FIG:2}
\end{figure}

PLMs provide assistance to urban researchers in tasks such as information retrieval, summarization, and tracking of urban institutions and documents. As the example shown in Figure \ref{FIG:2}, we utilized ChatGPT to obtain five institutional documents concerning urban land use. The PLMs effectively retrieved and cited the relevant documents through the assistance of the WebGPT plug-in.

\begin{figure}
	\centering
		\includegraphics[width=\linewidth, scale=0.75]{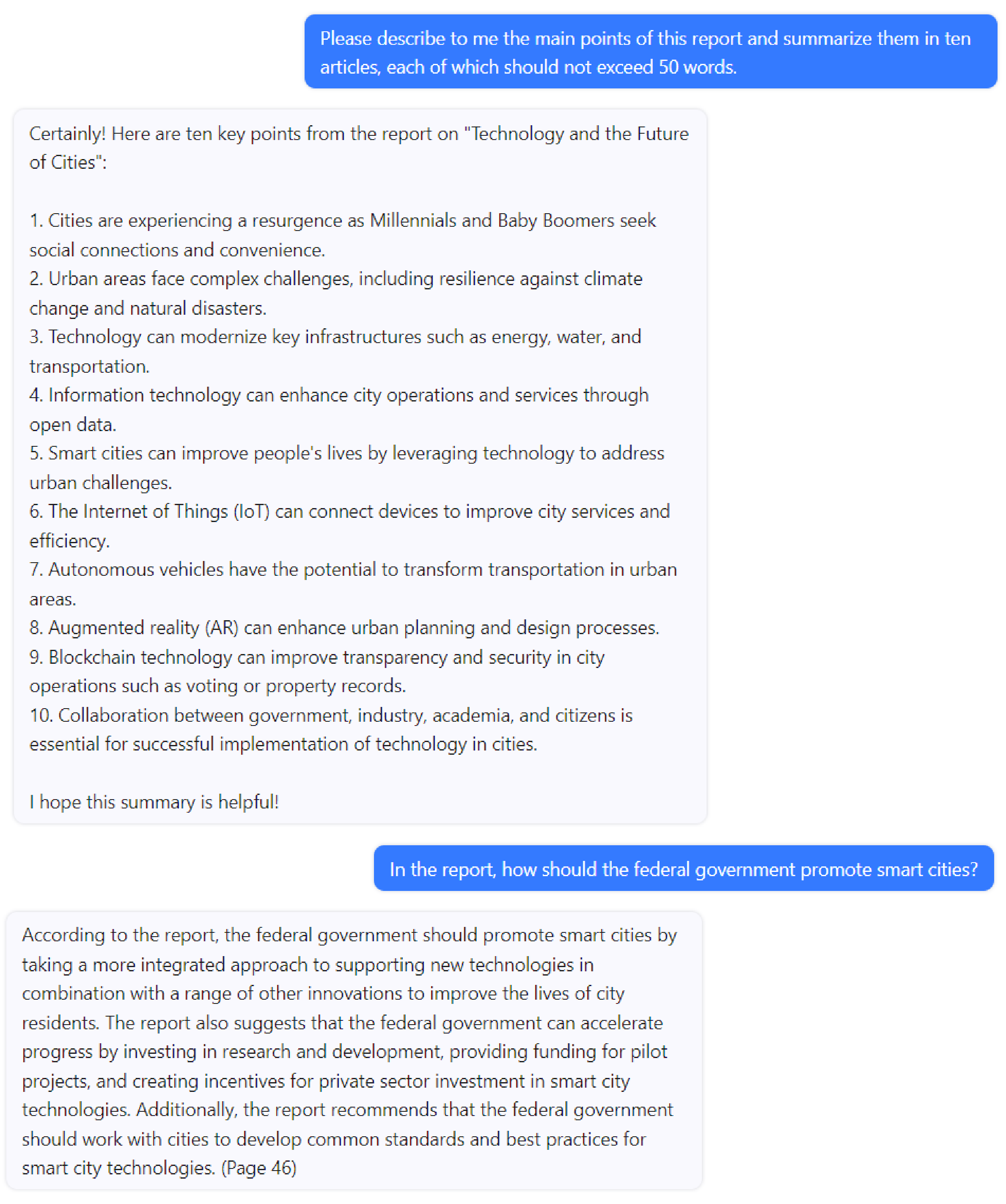}
	\caption{ChatPDF for summarizing documents (generated on May 2, 2023)}
	\label{FIG:3}
\end{figure}

In addition, PLMs possess remarkable capabilities for aggregation, allowing for the extraction of pertinent information from extensive materials and the automatic extraction of key points. As an illustration, we provided ChatPDF (an open tool based on the ChatGPT API) with a report by the President's Council of Advisors on Science and Technology (PCAST), Technology and the Future of Cities. We requested that ChatPDF extract and summarize the primary points of the document, as well as respond to specific inquiries regarding particular topics. ChatGPT successfully sorted and condensed the content in the report as requested. It also managed to locate and respond to the specific content in the institutional document while indicating the source of the answer (see Figure \ref{FIG:3}).

\begin{figure}
	\centering
		\includegraphics[width=\linewidth, scale=0.75]{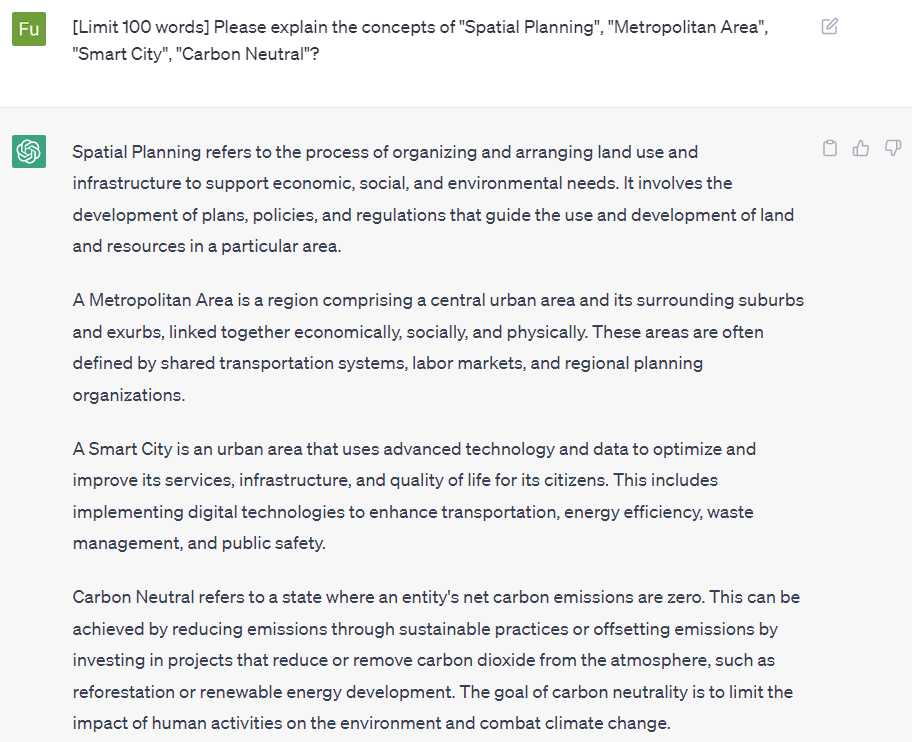}
	\caption{ChatGPT's feedback for explaining concepts (generated on May 2, 2023)}
	\label{FIG:4}
\end{figure}

PLMs are capable of acquiring and elucidating complex urban concepts, which can be particularly useful for researchers without a background in urban research. It can explain relevant terminology without requiring additional context. For instance, we tasked ChatGPT with explaining the meanings of various concepts that we had identified as relevant to urban research, such as "Spatial Planning", "Metropolitan Area", "Smart City", and "Carbon Neutral". As expected, ChatGPT was able to provide precise and accurate explanations of these concepts (see Figure \ref{FIG:4}).

\subsection{Urban space}
The study of urban space covers multiple dimensions such as geographic location, spatial form, spatial structure, land use, architectural form, and urban landscape \citep{koumetio2023artificial, sharifi2023progress}. These dimensions involve diverse textual and non-textual data sources. The advent of PLMs presents novel approaches for integrating multi-source data in urban space study. Due to its advanced natural language processing capabilities, ChatGPT is able to carry out tasks such as code generation and modification \citep{merow2023ai, sobania2023analysis}. This enhances the efficiency of urban spatial research by assisting in programming and streamlining the integration of novel data sources like cell phone signaling, points of interest (POI), and urban street views into urban spatial research. Two examples will be used to demonstrate the opportunities of PLMs in urban space study.

\begin{figure}
	\centering
		\includegraphics[width=\linewidth, scale=0.75]{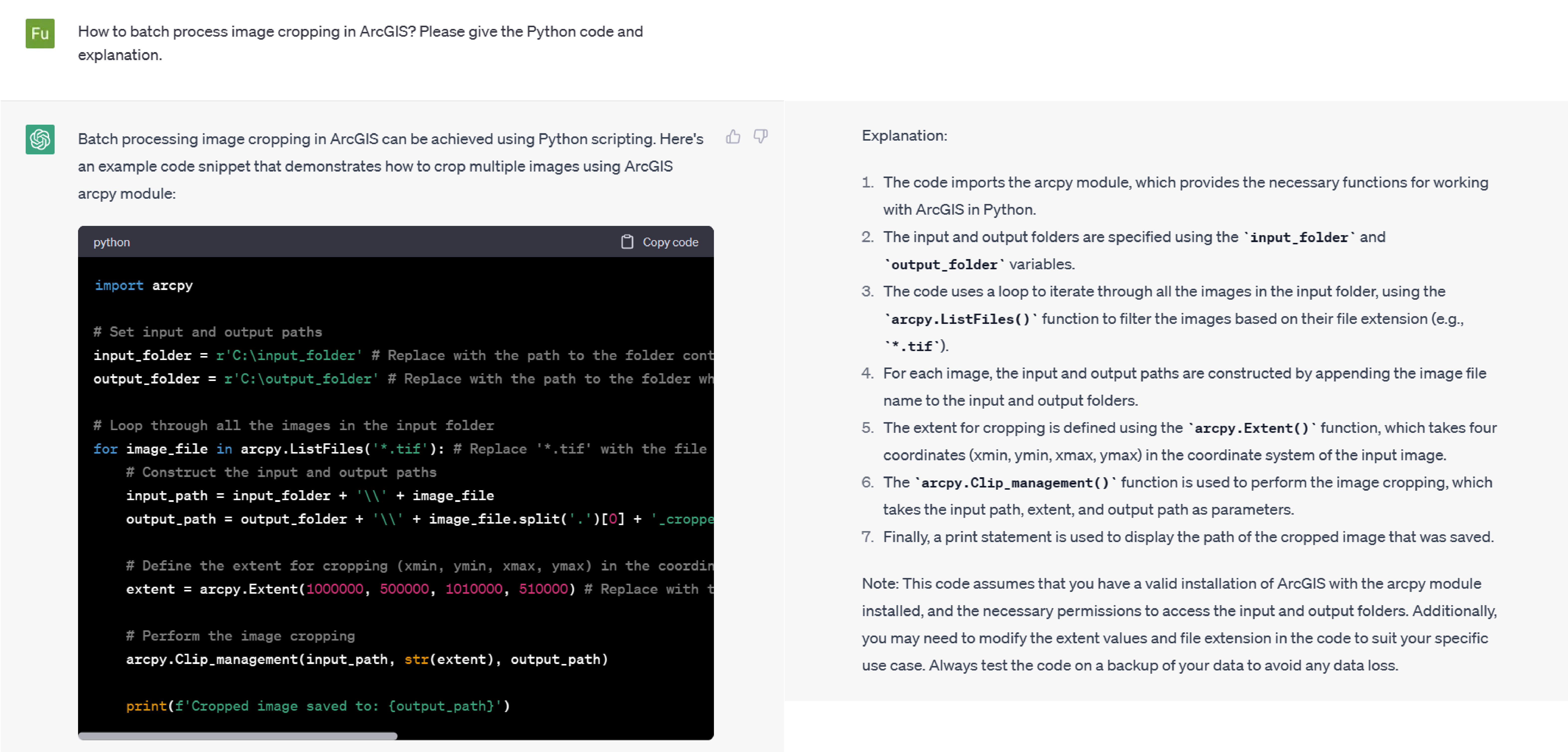}
	\caption{ChatGPT output code for ArcGIS operations (generated on April 17, 2023)}
	\label{FIG:5}
\end{figure}

\begin{figure}
	\centering
		\includegraphics[width=\linewidth, scale=0.75]{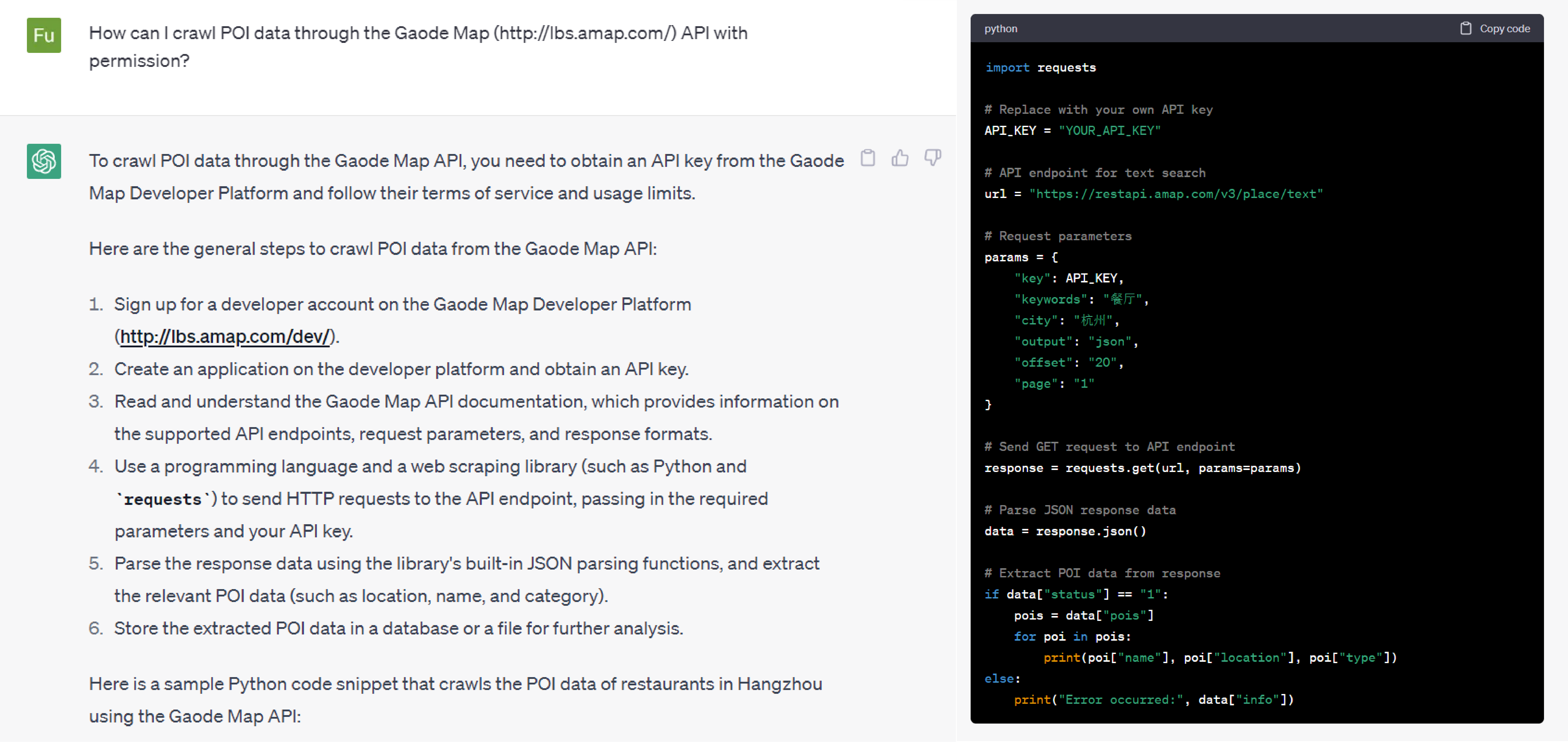}
	\caption{ChatGPT's feedback on POI acquisition (generated on April 23, 2023)}
	\label{FIG:6}
\end{figure}

\begin{figure}
	\centering
		\includegraphics[width=\linewidth, scale=0.75]{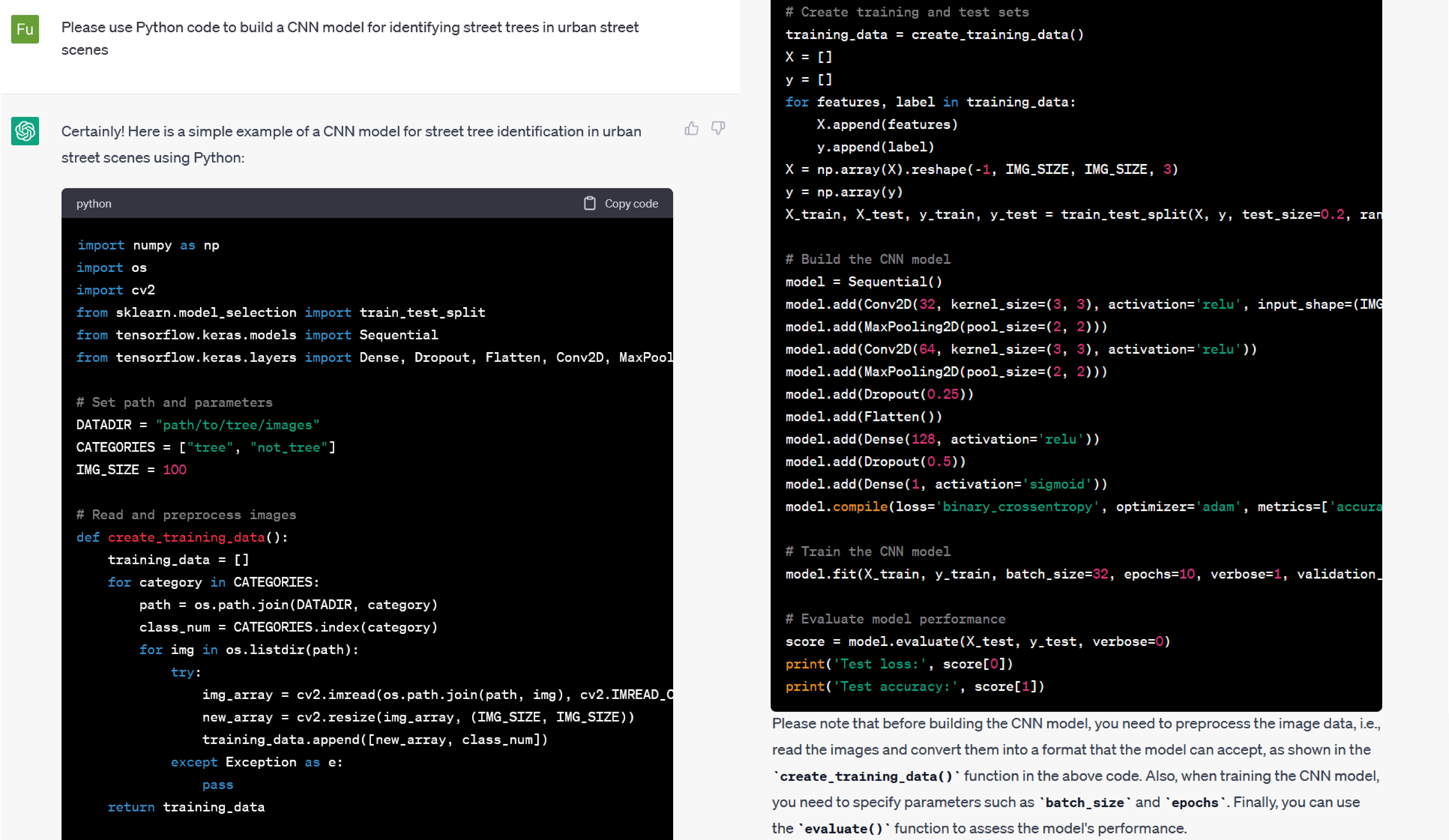}
	\caption{ChatGPT output code for street view recognition (generated on April 18, 2023)}
	\label{FIG:7}
\end{figure}

We used POI data to delineate central city boundaries, as an illustrative example. In this process, PLMs can assist with remote sensing imagery analysis, kernel density analysis, and other methods. For instance, we could query ChatGPT for guidance on "batch processing image cropping using ArcGIS, along with Python code and explanations," and employ ArcPy to execute the command (see Figure \ref{FIG:5}). PLMs can also assist with POI data crawling. We can make a request to ChatGPT: "How can we crawl POI data with permission through the AMap (\href{http://lbs.amap.com/}{http://lbs.amap.com/}) API?" ChatGPT can then provide the relevant code for POI data crawling (see Figure \ref{FIG:6}). PLMs have the ability to assist with programming, which makes them useful in urban streetscape recognition. One application of PLMs is to use models such as convolutional neural networks (CNNs) for urban landscape recognition. For instance, we utilized ChatGPT to help us construct a CNN model for the identification of street trees in urban streetscapes, using Python code (see Figure \ref{FIG:7}).

\subsection{Urban information}
Urban information refers to information generated by a multitude of data sources such as information and communication technologies (ICT), remote and physical sensors, and individuals \citep{wang2023defining}, and encompasses a wide range of topics including urban traffic, logistics, environment, disasters, and various types of urban economic information \citep{ismagilova2019smart}. PLMs can play a vital role in advancing urban information research. They can help identify geographic information in text using information extraction techniques, and facilitate monitoring and prediction of various urban phenomena such as disasters, housing prices, and traffic flow through the assistance of natural language processing, text mining, and machine learning. Here are two examples that illustrate the potential of PLMs in urban information research.

\begin{figure}
	\centering
		\includegraphics[width=\linewidth, scale=0.75]{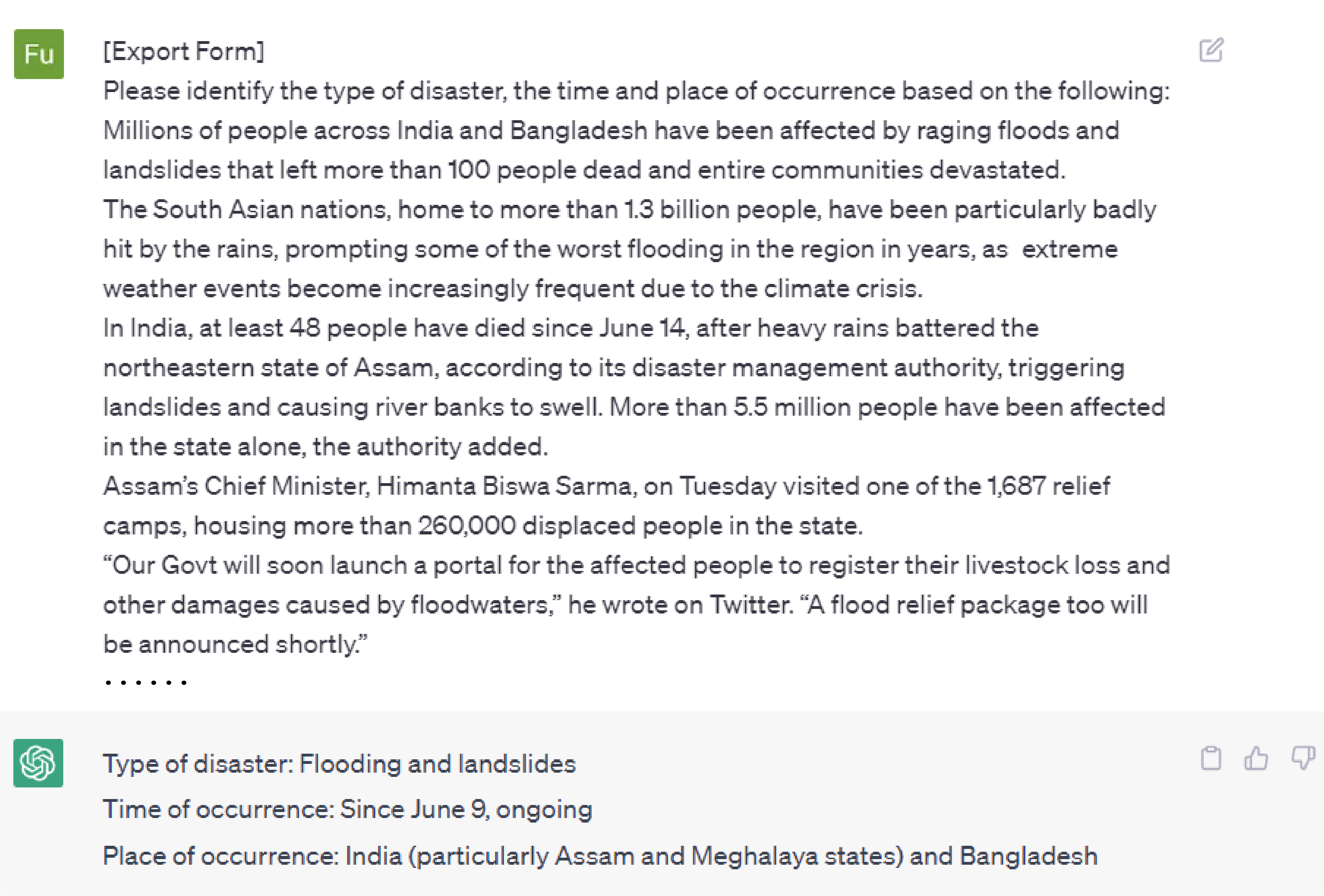}
	\caption{ChatGPT aids in disaster monitoring (generated on May 2, 2023)}
	\label{FIG:8}
\end{figure}

PLMs possess the capacity for helping monitor and predict natural disasters or public health events. Firstly, as an important function of PLMs, text mining has the ability to identify and extract disaster-related information from diverse sources, such as news articles, social media, and emergency reports. This information includes the time, location, and magnitude of the disaster. Secondly, the natural language reasoning capabilities of PLMs can aid in solving various comprehension and reasoning tasks, including scenario estimation for disaster monitoring and generating corresponding monitoring reports \citep{zheng2023chatgpt}. Additionally, time series analysis of disaster texts aids in achieving disaster prediction. As the example shown in Figure \ref{FIG:8}, we supplied ChatGPT with a text describing a disaster (extracted from a web report of the 2022 floods in Assam, India), and requested it to identify the time and location of the disaster and provide location details.

\begin{figure}
	\centering
		\includegraphics[width=\linewidth, scale=0.75]{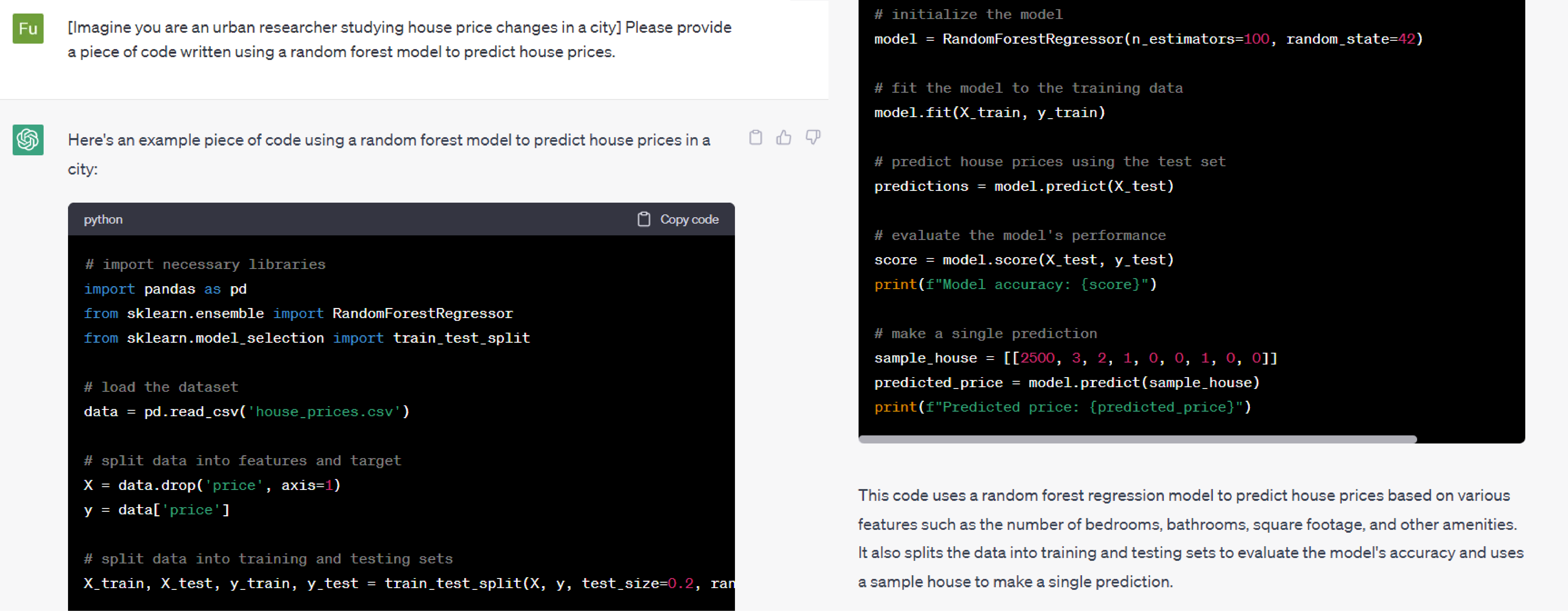}
	\caption{ChatGPT aids in predicting house prices (generated on May 2, 2023)}
	\label{FIG:9}
\end{figure}

PLMs are capable of assisting the forecasting of urban information, including housing prices, by utilizing various data sources such as demographic data, real estate listings, and local economic indicators. Moreover, we can perform data analysis to forecast future house prices in a particular area with the aid of auxiliary programming. As an illustration, we could request ChatGPT to construct a random forest model to predict the future trend of housing prices and provide us with the code for this prediction (see Figure \ref{FIG:9}).

\subsection{Citizen behaviors}
\begin{figure}
	\centering
		\includegraphics[width=\linewidth, scale=0.75]{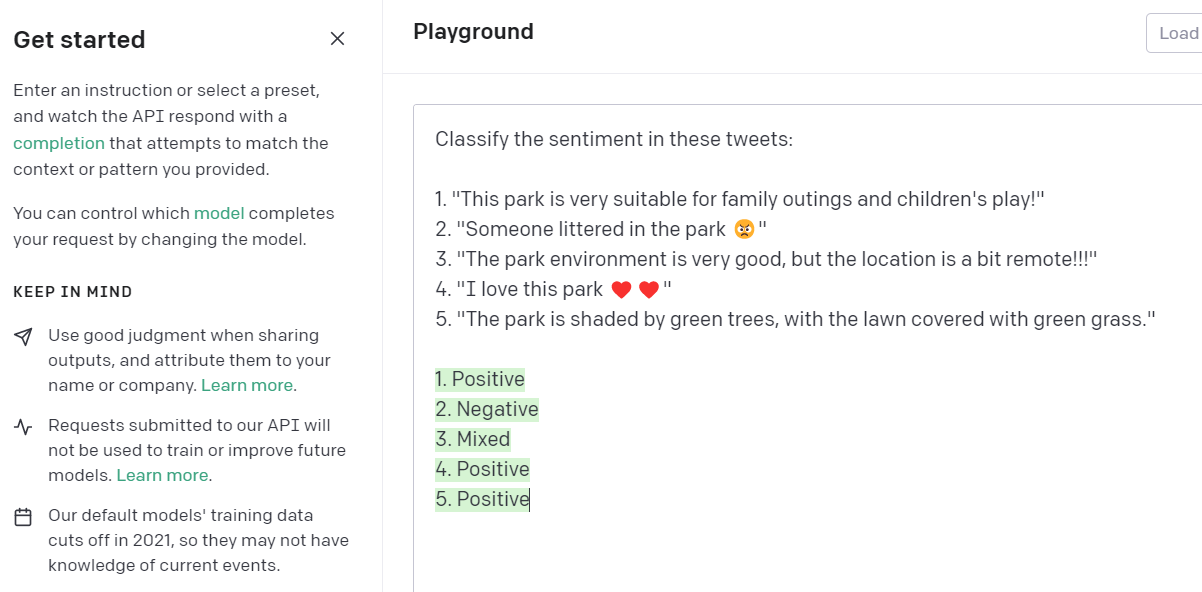}
	\caption{OpenAI API for sentiment analysis (generated on April 23, 2023)}
	\label{FIG:10}
\end{figure}

Research on citizen behaviors in cities covers issues such as public sentiment, population mobility, travel behavior, poverty and crime \citep{sharifi2023progress}. PLMs can help to study these issues. With their remarkable language processing capabilities, PLMs can parse social media texts, discern public events, track population movements, and monitor criminal activity, among other tasks. PLMs also possess powerful capabilities in sentiment analysis \citep{abdul2021framework} and gesture monitoring \citep{zhang2022would}. They can analyze the sentiment of posts, online comments, and various types of news or stories, and categorize them as either positive or negative, pros or cons. In addition, PLMs can analyze sentiment trends over time and detect significant changes in public opinion \citep{wang2023chatgpt}. This capability facilitates a comprehensive analysis of the shift in public sentiment towards an event or a location and encourages the utilization of social media in urban research \citep{abdul2021framework}. As an illustration, we presented ChatGPT with a set of paragraphs describing the utilization of the OpenAI API and its Tweet classifier to perform sentiment analysis on a comment regarding a certain park. ChatGPT was able to accurately identify the sentiment tendencies present in the comment (see Figure \ref{FIG:10}).

We summarize the possible applications of PLMs to urban institution, urban space, urban information, and citizen behaviors (see Table \ref{table3}).
\onecolumn
\begin{landscape}
\centering
    \begin{longtable}{p{1cm}p{3cm}p{4.5cm}p{7cm}p{1cm}p{2cm}p{2.5cm}}
    \caption{Summary of PLMs applications in urban research}
    \label{table3} \\   
        \hline
        Category & Topic & Process & Prompt examples & Data type & Data source & NLP type\\
        \hline
        \endfirsthead\\    
        \multicolumn{7}{c}{Table1: Summary of PLMs applications in urban research (continued)}\\
        \hline
        Category & Topic & Process & Prompt examples & Data type & Data source & NLP type\\
        \hline
        \endhead\\
        \hline
        \endfoot
        Urban institution & Retrieving and summarizing information on urban institution & Retrieving institutional documents, extracting and summarizing key information from the text & “What’s documents has the U.S. government issued for urban renewal?” or “Please help me organize the “The Strong Cities, Strong Communities (SC2) initiative” in item 4 above into five points.” & Text & System document & Information extraction, text summarization \\
        \quad & Explaining complex concepts & Explaining complex concepts in the urban systems literature & “Please explain to me the concepts of ‘spatial planning exploration’, ‘metropolitan symbiosis’, ‘digital planning’ and ‘carbon reduction planning’?” & Text & System document & Question answering \\
        \quad & Evaluating city system construction based on public feedback & Obtaining key messages from public comments and analyzing sentimental preferences for system building & “Please help me to extract the key information from the following feedback and analyze its emotional tendencies.” & Text & System document & Information extraction, sentiment analysis \\
        Urban space & Identifying geographic information & Identifying geographic information from text, including time, address, coordinates and other information & “Which landmark in China is described in the passage below? Where is this place (please indicate with latitude and longitude)?” & Text & Social media & Information extraction, question answering \\
        \quad & Identifying urban spatial patterns & Identifying city centers or city boundaries using POI data & “How can I crawl POI data through the AMap (http://lbs.amap.com/) API with permission?” & Spatial data & Points of interest (POI) & Assisted programming \\
        \quad & Identify urban spatial structure & Identifying urban spatial structure using smart-card data & "What are the algorithms for identifying urban spatial structure using community structure mining methods?" or "Can you provide a code example of community structure mining method to identify urban spatial structure?" & Spatio-temporal data & Smart-card data & Assisted programming \\
        \quad & Identify urban landscape & Building CNN model to identify urban streetscape & “Please use Python code to build a CNN model for identifying street trees in urban street scenes.” & Image & Street view image & Assisted programming \\
        \quad & Identifying urban land use change & Simulating and evaluating urban land use change using a metacellular automata model & "What data needs to be collected to assess urban land use change with metacellular automata?" or "Can you provide a code example of a meta-cellular automaton to assess urban land use change?" & Spatial data & Land use type & Assisted programming \\
        Urban information & Monitoring disaster & Identifying disaster-related information in Internet data, enabling disaster monitoring and prediction through keyword identification and modeling & "[Export Form] Please identify the type of disaster, the time and place of occurrence based on the following:"  & Text & Social media & Information extraction, assisted programming \\
        \quad & Predicting house price & Predicting house prices through economic data modeling & "How can a random forest model be used to predict house price changes when data such as house attributes are known?" & Number, image & Demographic data, house price, etc. & Assisted programming \\
        \quad & Suggesting infrastructure distribution & Suggesting distribution by identifying positive and negative public sentiment about infrastructure distribution & "Please help me analyze the emotional positives and negatives of each of the above comments." & Text & Social media & Sentiment analysis \\
        Citizen behaviors & Sensing public sentiment & Sensing the public's positive and negative emotions about a place or event & "Please analyze the emotional tendencies in each of the following statements:" & Text & Social media & Sentiment analysis \\
        \quad & Monitoring population movements & Identifying migration and movement of people in text, including time and place & "Please identify the main activities in the following reports, as well as the time and place." & Text & Social media & Information extraction \\
        \quad & 1Monitoring crime & Identifying suspected criminal information and negative information in social media & "Please identify whether there is suspected criminal information or negative information in the text, and if so, identify the specific content." & Text & Social media & Information extraction \\
    \end{longtable}
\end{landscape}

\twocolumn

\section{Challenges}
\subsection{Technical perspective}
\subsubsection{Technical restrictions}
\begin{figure}
	\centering
		\includegraphics[width=\linewidth, scale=0.75]{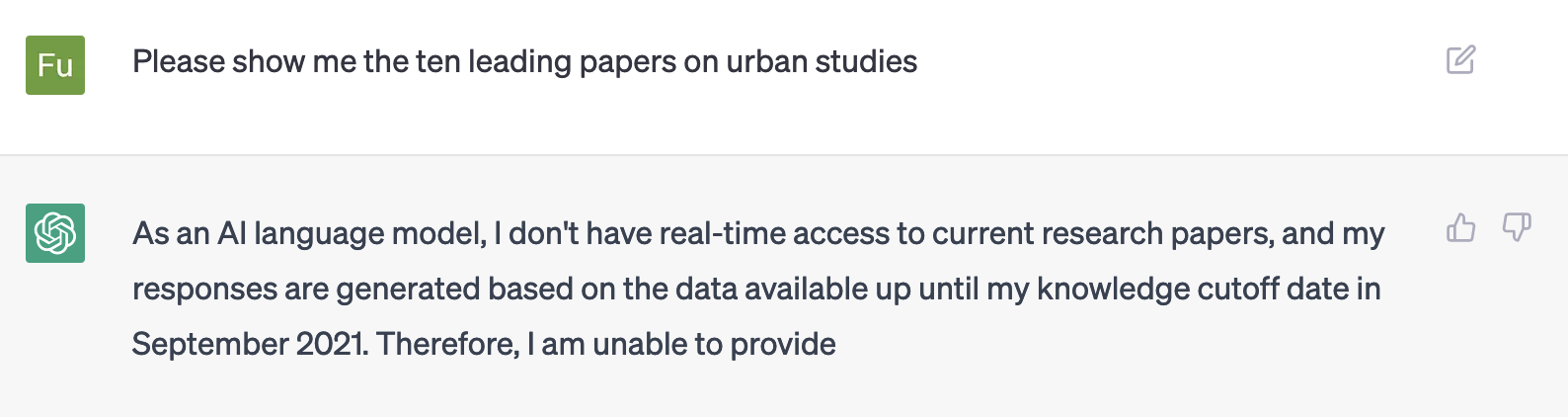}
	\caption{ChatGPT's feedback for providing papers (generated on April 16, 2023)}
	\label{FIG:11}
 
\end{figure}

Time restrictions: PLMs require vast amounts of data to be trained for initial models. For instance, ChatGPT's training data only goes up to June 2021 \citep{zhu2023chatgpt}. This means that ChatGPT can only understand and infer information from 2021 and earlier, making it challenging to update to the most current data \citep{teubner2023welcome}. Therefore, when asked to provide ten authoritative papers on urban research, ChatGPT was unable to provide current research papers in real time due to data training time constraints (see Figure \ref{FIG:11}).

\begin{figure}
	\centering
		\includegraphics[width=\linewidth, scale=0.75]{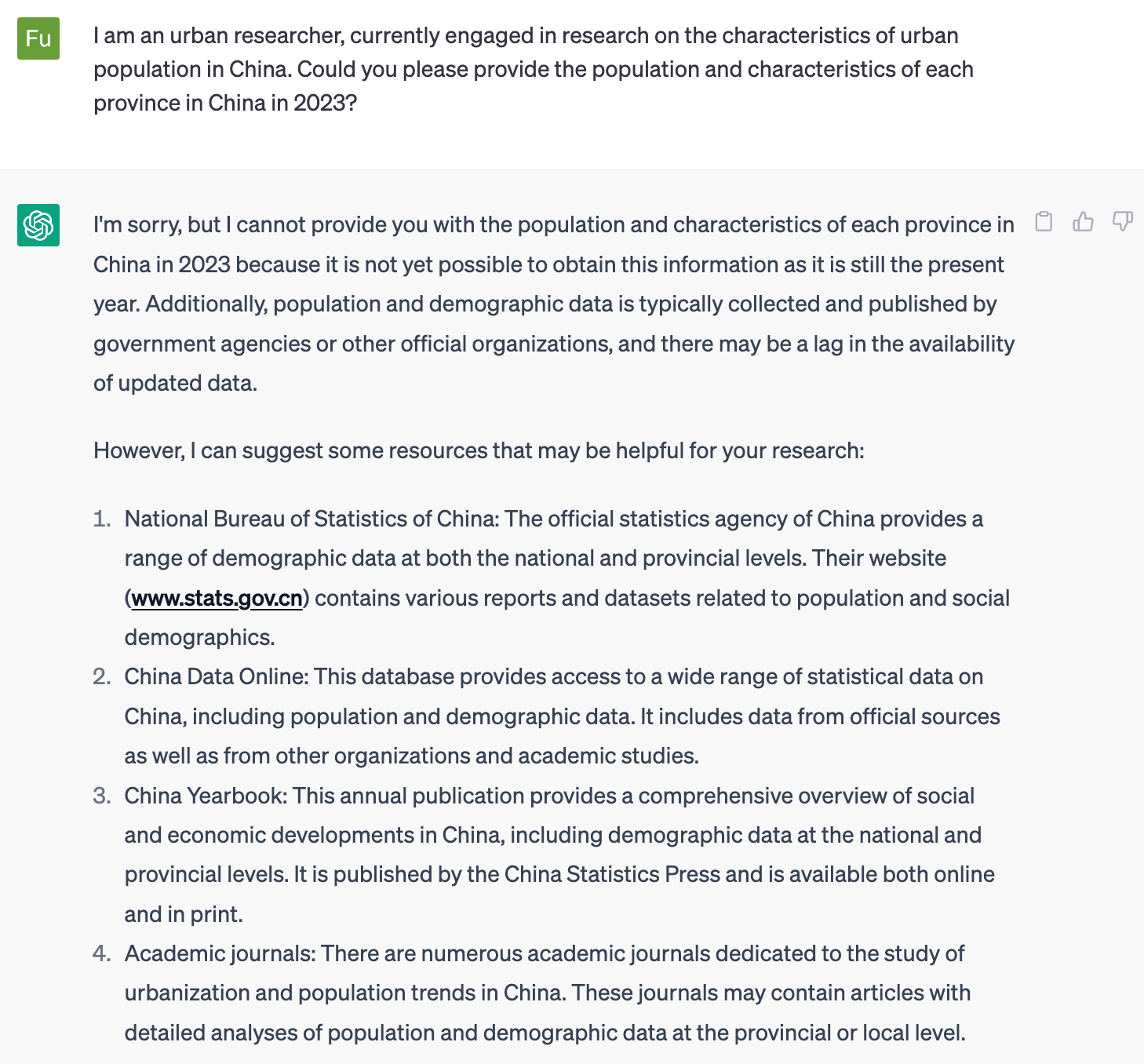}
	\caption{ChatGPT's feedback on data collection (generated on April 22, 2023)}
	\label{FIG:12}
\end{figure}

Permission restrictions: The issue of data restrictions in PLMs is further exacerbated by the incompleteness and inaccessibility of big data \citep{salganik2019bit}. Although ChatGPT is capable of searching networks and providing citation source annotations after using the WebGPT plugin, there are still significant limitations in data collection, such as inaccessible cell phone signaling and travel data. This hinders researchers from using PLMs to obtain authoritative information for urban studies. As an example shown in Figure \ref{FIG:12}, when trying to study urban demographic characteristics, we attempted to ask ChatGPT about the current demographic characteristics of each province in China. However, ChatGPT indicated that it was unavailable due to training data limitations and access restrictions. This indicates that researchers still need to manually retrieve data from specialized databases instead of relying solely on PLMs for conducting research on recent data.


Modality restrictions: Presently, the multimodality of PLMs is mainly exhibited in the inference and analysis of data and text. For other modes like images, audio, and video, plug-ins and auxiliary programming are often required \citep{yang2023harnessing}. It is also challenging for PLMs to directly recognize remote sensing images in urban research, and it is difficult to conduct application in urban soundscape and urban images.

\subsubsection{Authenticity and validity}
\begin{figure}
	\centering
		\includegraphics[width=\linewidth, scale=0.75]{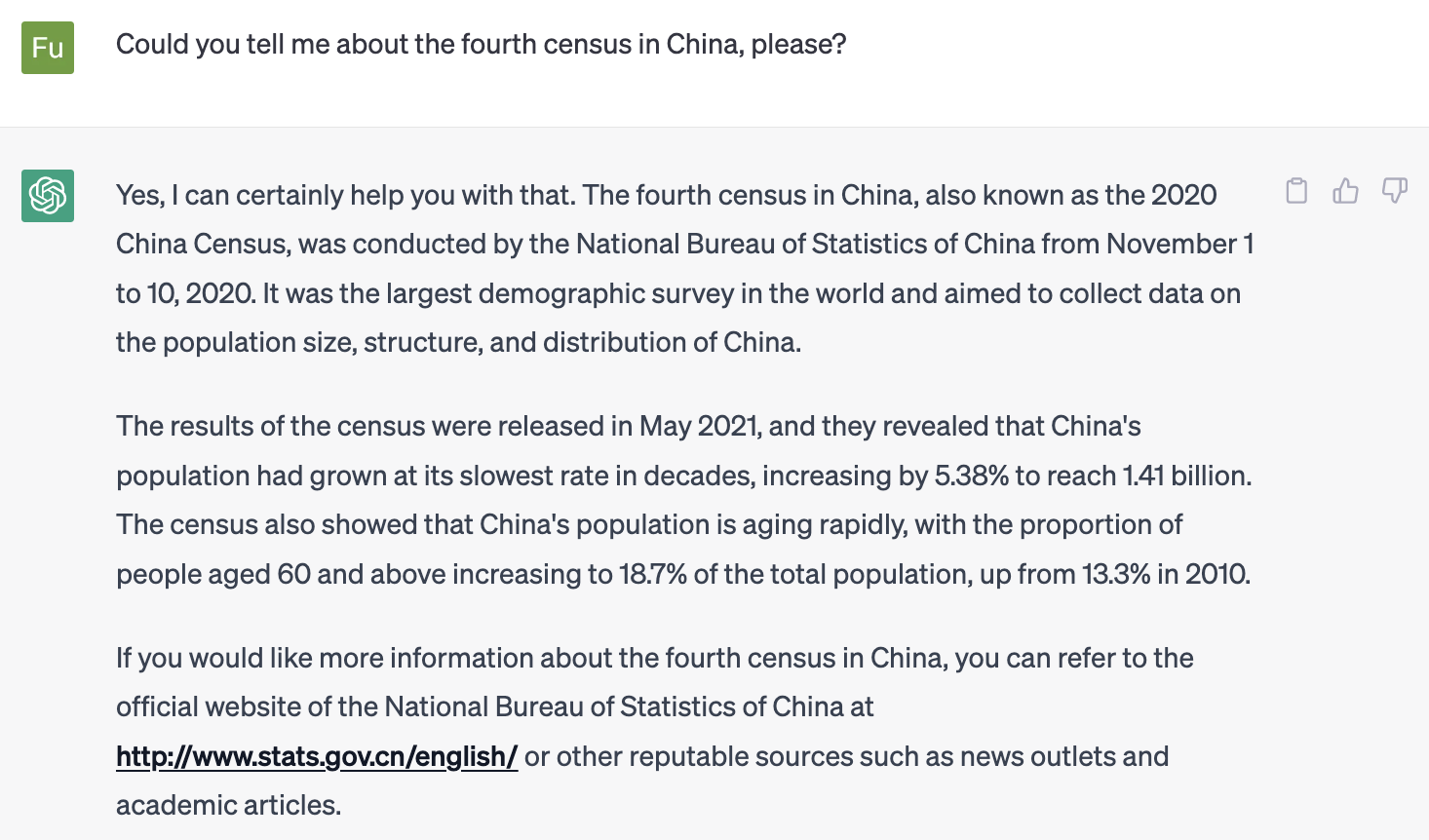}
	\caption{ChatGPT's feedback on data retrieval (generated on April 23, 2023)}
	\label{FIG:13}
\end{figure}

On one hand, it is worth noting that PLMs may generate false information, including fabricated literature \citep{haluza2023artificial} and factual errors (hallucinations) \citep{wu2023brief}, particularly in low-resource settings. On the other hand, the performance of PLMs is not uniformly consistent and stable, which may result in disparate responses to the same query \citep{liu2023summary}. For instance, when we inquired about "information on the fourth census of China", ChatGPT provided wholly inconsistent data, which could lead to entirely erroneous conclusions in urban studies (see Figure \ref{FIG:13}). It can be observed that ChatGPT does not currently offer a precise and dependable source of information for urban research, nor does it have the capability to effectively integrate diverse types of knowledge. To ensure the reliability and accuracy of PLMs' output information, particularly regarding issues concerning temporal and numerical dimensions like urban time series change and population change, a more rigorous validation approach is necessary.

\subsubsection{Comprehension skills}
PLMs, being a form of artificial intelligence, are essentially based on inferences about statistical relationships and currently lack the higher-order thinking skills to understand context and nuance \citep{liu2023summary}. In the context of complex urban research, this shortcoming can lead to the production of inaccurate data and misinterpretations \citep{kooli2023chatbots}, resulting in responses that lack depth and insight or even deviate from the intended topic \citep{farrokhnia2023swot}.

\begin{figure}
	\centering
		\includegraphics[width=\linewidth, scale=0.75]{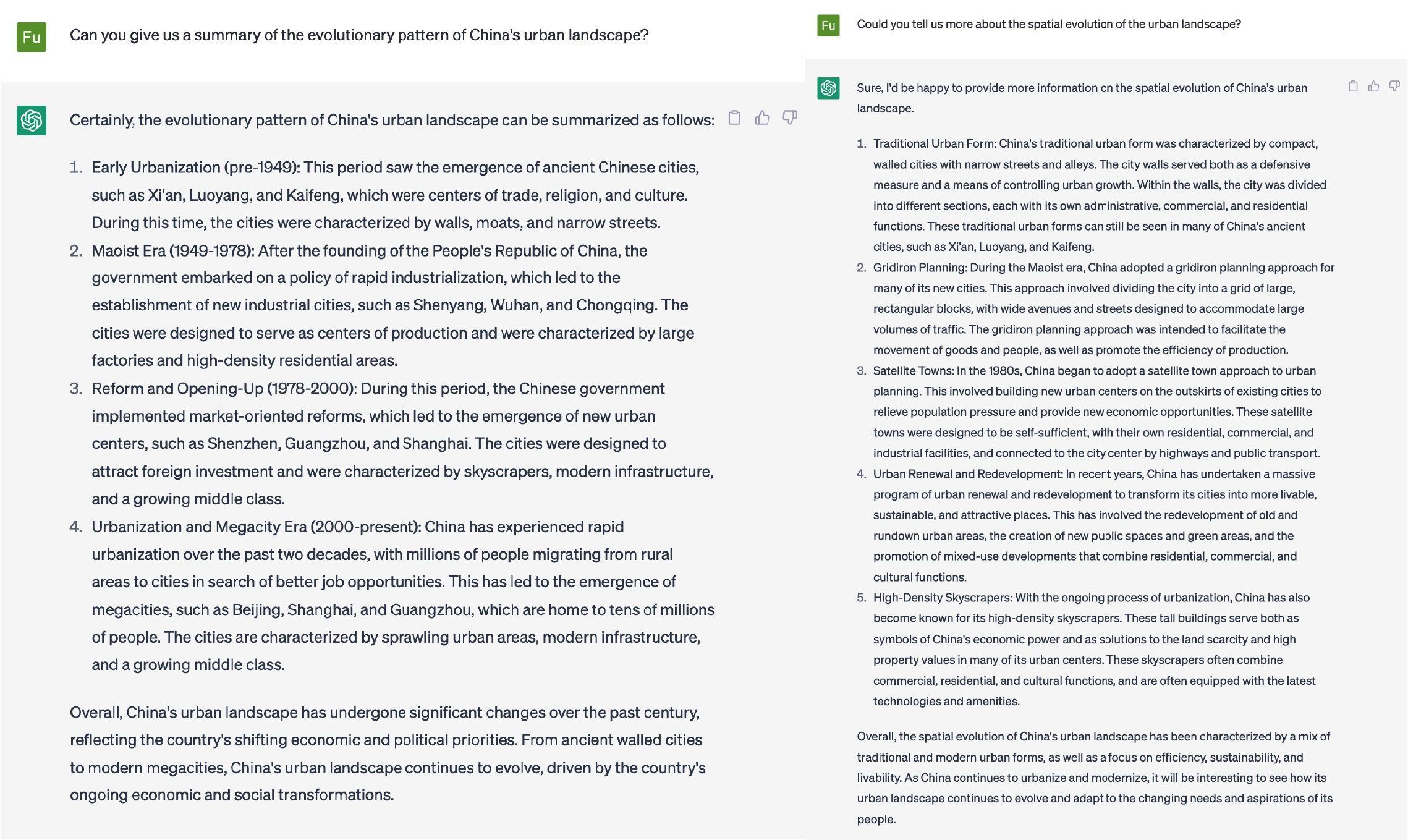}
	\caption{ChatGPT's feedback for summarizing patterns (generated on April 23, 2023)}
	\label{FIG:14}
\end{figure}

Furthermore, most PTMs are trained using general-purpose data sources, such as Wikipedia, which can limit their effectiveness in specific domains \citep{qiu2020pre}. For instance, when prompted to provide information on the "evolutionary patterns of Chinese landscape", ChatGPT could only offer superficial observations, struggling to grasp the underlying evolutionary patterns (see Figure \ref{FIG:14}). Consequently, generic PLMs continue to face limitations in comprehending intricate urban theories or patterns. While there exist PLMs that specialize in geography, such as ERNIE-GeoL, GeoBERT, and SpaBERT, their current use in the field of urban research is restricted by permissions and limited functionality, such as the classification and matching of POI, address segmentation, and geographic entity coding.

\subsection{Sociological perspective}
\subsubsection{Lack of trust}
The technical black box is an important feature of AI development \citep{yigitcanlar2020sustainability}. The PLMs, such as ChatGPT, are capable of providing feedback to users, yet they are incapable of elucidating the computational process that underlies their decision-making and predictive capabilities \citep{sanderson2023gpt}. This limitation leads to a dilemma in the application of PLMs in urban research. On the one hand, PLMs cannot guarantee the source and reference of generated information. The opacity of PLMs could potentially result in significant consequences when dealing with certain NLP tasks that demand high precision in the context of urban research. On the other hand, the public, who is one of the focal groups of urban research, may not have confidence that their private information is not being utilized for data retrieval and processing, thereby undermining public trust in PLMs. Consequently, PLMs need to augment their transparency and traceability, through algorithmic optimization or legal regulations, to address the expectations of both researchers and the public.

\subsubsection{Social bias and discrimination}
The training data for PLMs is typically obtained from publicly available web resources. However, there exists a significant amount of biased data on the internet, including information related to race, religion, and gender, among others \citep{buolamwini2018gender}. This bias can persist and be reflected in PLMs after training \citep{farrokhnia2023swot, jungwirth2023artificial}. Such biases in the model can have a harmful impact on the relevant groups of the public, perpetuating stereotypes and derogatory images \citep{brown2020language}. Furthermore, the population that utilizes internet resources has certain group characteristics, resulting in training samples that are biased and fail to accurately reflect the requirements of marginalized groups \citep{yang2023harnessing}. And one of the purposes of urban research is to promote equal and sustainable development of urban citizens \citep{meerow2019social}. Discrimination and prejudice can have significant social harm, even with minor deviations \citep{liang2022holistic}, resulting in unreasonable allocation of urban space, unjust public decision-making, and widening urban-rural divide.

\subsubsection{Threat to information safety}
Security and privacy are key issues to consider in the applications of PLMs to urban research. By virtue of utilizing researchers' queries and input data as their training material \citep{clarke2023call}, PLMs may potentially give rise to issues of data leakage and data theft. Such circumstances can result in data leakages of individuals and cities, thereby threatening personal privacy and city security. PLMs, such as ChatGPT, is possible to steal personal information from cities or the public through phishing emails and malware \citep{wu2023brief}, thus threatening city security and personal privacy. Furthermore, trained data by PLMs may be biased or erroneous, potentially yielding harmful output. PLMs are highly communicative and interactive. If harmful content is disseminated in large quantities, it can trigger a serious "infodemic" phenomenon \citep{de2023chatgpt, zarocostas2020fight}, generating mass anxiety, hate speech, and even urban riots, thereby jeopardizing urban public safety. Consequently, researchers should be circumspect with respect to sensitive information provided to PLMs, while simultaneously considering the security of PLMs' answers and strengthening the safety of urban and personal information.

\section{Future directions}
Based on the aforementioned exploration of the opportunities and challenges surrounding PLMs applied to urban research, we put forward several potential avenues that can enhance the role of PLMs in urban science research:

First of all, fundamental models based on urban research areas can be developed. As a consequence of the requirement for extensive model multimodal applications in urban research, coupled with the restrictions on using current models, the development of fundamental models within the realm of urban research could emerge as a novel avenue \citep{wang2023chatgpt}. This approach would incorporate multimodal applications, such as text, data, image, audio, and video, to extend the utilization of multi-source big data in urban research. Also, the foundational models customized for urban research could enhance the accuracy and precision of results, facilitating more intricate urban research tasks, such as the exploration of complex urban theories and laws.

Secondly, human-AI collaboration can be applied to facilitate urban research. The text analysis, abstract summarization, and assisted programming capabilities of PLMs have the potential to significantly enhance the research efficiency of urban researchers. PLMs can help strengthen the academic exchange of urban research and enhance the diversity of perspectives \citep{van2023chatgpt}. Furthermore, the integration of PLMs with emerging techniques such as deep learning can aid researchers in overcoming technical limitations and adapting to new urban research methods in the context of big data. This, in turn, would allow researchers to focus more on urban theoretical research and paradigm innovation. Finally, PLMs are expected to provide technical support for new directions in urban research, such as digital twin cities.

Thirdly, PLMs can be used to improve public participation and urban decision-making. On one hand, PLMs, such as ChatGPT, possess natural language interaction capabilities, which can be utilized to disseminate urban information to the public, thus advancing their comprehension and participation in urban research \citep{casares2018brain}. PLMs are also expected to promote urban research by understanding cities from a more human perspective through deep learning of public opinions. On the other hand, PLMs are poised to provide crucial assistance for urban decision-making, mitigating the undue impact of subjective factors on urban decision-making, and proposing ideas for the optimization of urban decision-making.

Finally, there is a need to be wary of falsehood, privacy, and liability issues. As previously mentioned, issues such as limited data, falsity, and social bias are major concerns that need to be addressed. However, there is no clear consensus on how ChatGPT can regulate these issues related to accuracy, privacy, and liability. As such, it is important to exercise caution and skepticism when using PLMs, to improve our judgment on PLMs answers and to view them as tools rather than relying on them completely \citep{krugel2023chatgpt}.

\section{Conclusion}
In this paper, we discuss the opportunities and challenges of PLMs in urban science research, using ChatGPT as an example. PLMs play a crucial role in the study of urban institution, urban space, urban information, and citizen behaviors. The benefits of PLMs in question answering, abstract summarization, and analysis enhance text retrieval efficacy and facilitate the explication of intricate concepts in institutional documents. PLMs can facilitate the applications of new technologies and data in urban research, including through assisted programming. Additionally, the strengths of PLMs in information extraction and text classification enable text-based data to be utilized in urban research, amplify the availability of big data sources for cities, and supply new insights for urban research.

Nevertheless, PLMs still confront numerous challenges in urban research. The issues of temporal limitation, authoritative limitation, modality limitation, credibility, and weak comprehension have been exposed in studies and still pose multiple challenges. Public trust, social biases, and public safety represent significant limitations to the practical applications of PLMs in urban research. These issues require further discussion and consideration.

PLMs will become a potent instrument for urban researchers. We hope to further promote the applications of PLMs in urban research by developing FMs based on the field of urban research, in order to enhance the applications of new urban research and practice in the context of big data.

\printcredits

\bibliographystyle{cas-model2-names}

\bibliography{cas-refs}

\end{document}